\documentclass[12pt]{article}
\usepackage{graphicx,amsmath,amssymb,multirow}

\usepackage{color}
\definecolor{text1}{rgb}{0,0,1} 
\definecolor{text2}{cmyk}{0,.71,0.88,0.17} 
\definecolor{text3}{cmyk}{0,0,0,1} 
\definecolor{text4}{rgb}{0,1,0} 
\definecolor{text5}{rgb}{1,0,0.5} 
\definecolor{text6}{rgb}{1,0,0} 
\newcommand{\mcj}[1]{{\color{text3}#1}}

\newcommand{\red}[1]{{\color{text3}#1}}
\newcommand{\ddd}[1]{{\color{text3}#1}}
\newcommand{\mage}[1]{{\color{text3}#1}}
\newcommand{\kevin}[1]{{\color{text3}#1}}

\begin{document}

\parindent=0pt

\def\overset{\stackrel{1-1}{\leftrightarrow}}

\begin{center}

{\Large {\bf A Bivariate Power Generalized Weibull}}
\medskip

{\Large {\bf Distribution: a Flexible Parametric}}\medskip

{\Large {\bf  Model for Survival Analysis}}
\bigskip

{\large by \textsl{M.C. Jones, \ddd{Angela Noufaily}} and
\textsl{\ddd{Kevin Burke}}}
\end{center}

\bigskip
\red{
\noindent \textbf{Abstract}\medskip

We are concerned with the flexible parametric analysis of bivariate survival data.
Elsewhere, we have extolled the virtues of the ``power generalized Weibull'' (PGW)
distribution as an attractive vehicle for univariate parametric survival analysis: it
is a tractable, parsimonious, model which interpretably allows for a wide variety of
hazard shapes and, when adapted (to give an adapted PGW, or APGW, distribution),
covers a wide variety of important special/limiting cases. Here, we additionally
observe a frailty relationship between a PGW distribution with one value of the
parameter which controls distributional choice within the family and a PGW
distribution with a smaller value of the same parameter. We exploit this frailty
relationship to propose a bivariate shared frailty model with PGW marginal
distributions: these marginals turn out to be linked by the so-called BB9 or ``power
variance function'' copula. This particular choice of copula is, therefore, a natural
one in the current context. We then adapt
the bivariate PGW distribution, in turn, to
accommodate APGW marginals.  We provide a number of
theoretical properties of the
bivariate PGW and APGW models and show the potential of the latter for practical work
via an illustrative example involving a well-known retinopathy dataset, for which the
analysis proves to be straightforward to implement and informative in its outcomes.
The novelty in this article is in the
appropriate combination of specific
ingredients into a coherent and successful whole.
\bigskip

\noindent \textit{Keywords}: BB9 copula; Gompertz; log-logistic; power variance frailty; shared frailty.}
\medskip

\bigskip
\newpage

\noindent \textbf{1.~Introduction}\medskip

In this article, we are concerned with the flexible parametric
analysis of
paired survival data. Such data arise frequently in medicine, \ddd{for
example,}
when comparing treatment and control on pairs of related sampling units
such as an individual's eyes \ddd{or limbs, or when measurements are made
pre- and
post-intervention of some kind, or when familial data such as observations
made on twins or on a parent and child are of interest,
and so on. Here, our concern is with} providing a
flexible
parametric model for the entire, correlated,
bivariate distribution of the pairs of outcomes.
To do so, we reason for specific  elements and combine them into a
new overall model in a coherent and successful way.
\medskip

In Burke, Jones \& Noufaily (2018; henceforth BJN), we argued, in the
univariate
case, in favour of
flexible parametric survival analysis in general and for the
use of an
adapted form of \mage{an existing} flexible parametric model called the ``power
generalized
Weibull'' (PGW) distribution in particular.
Advantages of the latter include that its two shape parameters control
key shapes of the hazard function (constant, increasing, decreasing,
up-then-down,  down-then-up, and no others) and that, when adapted,
several common and
important survival distributions are special/limiting cases of it
(log-logistic, Weibull, Gompertz and others). \red{The PGW distribution is
one of only a very few to parsimoniously and interpretably control hazard
shapes as just described | others include the generalized gamma (GG) and
exponentiated
Weibull (EW) distributions | but it is preferred by us because of its
extra
tractability
and its greater breadth of particular cases.
\medskip

In this article, we also take advantage of a further feature of the PGW
distribution not shared with GG or EW distributions.}
Inter alia, in Section \mcj{2.1, we obtain} a frailty relationship
between a PGW distribution with one value of the parameter
$\kappa$ which controls specific distributional choice within the
 PGW family and a smaller value of the same parameter.
\mcj{We then} exploit this frailty relationship and first pursue --- in
Section 3 --- a natural
extension of the
PGW distribution,  to the multivariate
case in principle, but more specifically, for convenience and many
practical applications, to the bivariate case,
through the shared frailty route.  In Section 4, we build on the work of Section 3 to
provide a closely related bivariate distribution with adapted PGW (APGW)
marginals,
which is the version that we  suggest for practical work, as exemplified
in Section 5. \red{The distribution has eight basic parameters, and these can be
extended in natural ways
to accommodate covariates. Even so, the model proves to be straightforward to
implement using maximum likelihood techniques, and to provide interpretable and
insightful analysis of the example on which we illustrate the methodology.}
\medskip

In more detail, we first provide the relevant univariate background in Section
2\mcj{, including the univariate frailty result that drives the remainder
of the paper in Section 2.1}. The bivariate
PGW distribution of interest in this article is \mcj{then} developed in
Section
3.1: its conditional hazard functions are considered in Section 3.2, its cross ratio
dependence function in Section 3.3, and its copula representation, and properties
ensuing therefrom, in Section 3.4.
After explaining why we cannot follow the same shared frailty approach for
the APGW distribution as we do for the PGW distribution, we propose  a
bivariate APGW distribution by transforming PGW marginal
distributions to APGW marginal distributions, retaining the same
copula; see
Section 4.1. Properties other than those solely dependent on the copula,
which are the same for both bivariate PGW and APGW distributions,
are considered in Section 4.2. We provide an illustrative example of the
application of the APGW distribution to a standard, retinopathy, dataset
in Section 5, first without its covariate (Section 5.1) and then with the
inclusion of  the covariate (Section 5.2).
\mage{In particular, we observe how much
the choice of marginal distributions impacts on the degree of dependence of the
bivariate
failure model.} We conclude the article with brief further discussion in Section
6.\medskip

As the  reader will observe, the distributions of interest in this article
\red{retain the} high
degree of tractability \red{of their univariate counterparts}.

\bigskip
\medskip
\noindent \textbf{2.~Univariate background}\medskip

Write PGW$(\gamma,\kappa,\lambda)$ for
the PGW
distribution with power parameter $\gamma>0$,
distribution-choosing
parameter $\kappa>0$ and vertical scale/proportional hazards (PH)
parameter $\lambda>0$.
 That is, it has cumulative hazard function (c.h.f.)
$$\lambda\,H_N(t; \gamma,\kappa) = \lambda
\left\{(1+t^\gamma)^\kappa-1\right\}.$$
  In practice, it is also important to
consider a horizontal scale/accelerated failure time (AFT)
parameter $\phi>0$ which enters the c.h.f.\ via $\lambda H_N(\phi t;
\gamma,\kappa)$; for theoretical purposes, we can set $\phi=1$ without
loss
 of generality. \red{The PGW distribution was first introduced by Bagdonavi\c{c}ius
\& Nikulin
(2002) (see also Nikulin \& Haghighi, 2009), independently re-introduced by
Dimitrakopoulou, Adamidis \& Loukas (2007), and recognised as an interesting
competitor to the GG and EW distributions in Jones \& Noufaily (2015).}\medskip

The shape parameters $\gamma$ and $\kappa$
control the `head' (values near
zero) and tail of the distribution in the sense that the
hazard function $h_N(t;\gamma,\kappa) = H'_N(t;\gamma,\kappa)$
behaves as $t^{\gamma-1}$ as $t \to 0$ and as $t^{\gamma\kappa-1}$
as $t \to\infty$. This allows a hazard function with a zero,
finite or infinite value at its head and likewise, independently, at its
tail.
What is more, the hazard function joins head to tail in a smooth
manner
which yields a decreasing hazard when $\gamma \leq 1$ and  $\kappa\gamma
\leq 1$, an increasing hazard when $\gamma \geq 1$ and  $\kappa\gamma
\geq 1$, an up-then-down (often called `bathtub')  hazard when
$\gamma \geq 1$ and  $\kappa\gamma
\leq 1$, and a down-then-up (sometimes called `upside-down bathtub')
 hazard when
$\gamma \leq 1$ and  $\kappa\gamma
\geq 1$.
If $\gamma=\kappa=1$, the hazard function is constant (the PGW
distribution is then the exponential distribution).
\medskip

BJN principally work with an adapted form of the PGW distribution,
written APGW$(\gamma,\kappa,\lambda)$; this has as its c.h.f.\
a horizontally and
vertically rescaled form of the basic PGW c.h.f.:
$$\lambda\,H_A(t; \gamma,\kappa) = \lambda
\,\left(\frac{\kappa+1}\kappa\right)
\left\{\left(1+\frac{t^\gamma}{\kappa+1}\right)^\kappa-1\right\}.$$
Here, $\gamma, \lambda>0$ and $\phi=1$ as before, but the domain of
$\kappa$ can be extended (if desired) from $\kappa>0$ to $\kappa>-1$; this
affords a cure model when $-1<\kappa<0$, in which  the `improper' survival
function tends to $\exp\{\lambda(\kappa+1)/\kappa \}$ as $t \to \infty$.
Of course, when $\kappa>0$, hazard function shapes, as described above for
$h_N$,
are unaffected.
\medskip

When $\kappa=1$, (A)PGW distributions are Weibull distributions; when
$\kappa=2,$\linebreak $\gamma=1$, they are linear hazard distributions.
The reason for switching from $H_N$  to $H_A$ is that
$H_A$ readily accommodates limiting cases, which also turn out to be
important
and popular survival models: $\kappa=0$ corresponds to the Burr Type XII
distribution which incorporates the log-logistic distribution when
$\lambda=1$; and when $\kappa \to \infty$, the PGW distribution tends to
a form sometimes called the `Weibull extension' model which, for
$\gamma=1$, is the  Gompertz distribution. The APGW distribution therefore
affords, by choice of $\kappa$, a wide range of popular survival models,
from
light-tailed Gompertz, through the ubiquitous Weibull, to the
heavy-tailed log-logistic, and `beyond' to certain cure models.
\medskip\medskip

\noindent \textsl{\mcj{2.1~Frailty links}}
\medskip

Frailty is usually introduced into survival models by mixing over
the
distribution of the proportionality parameter \mcj{$B$ say
(Hougaard, 2000, Duchateau \& Janssen, 2008, Wienke, 2011)}. A given
survival distribution can be produced  from another given
survival distribution by such frailty mixing if the ratio of their hazard
functions is decreasing
(\mcj{Gupta \& Gupta, 1996}). Thus, when  PGW$(\mcj{\gamma,\, \kappa,\,
B})$
is mixed with a certain frailty
distribution, PGW$(\mcj{\gamma,\, \omega \kappa,\, \lambda})$ for $0 \leq
\omega \leq 1$ and appropriate \mcj{$\lambda>0$} results.
\mcj{Notice that the frailty mixing `moves us down' from a PGW
distribution with
parameter $\kappa$ to a PGW distribution with (the same value of $\gamma$
and) smaller parameter $\omega \kappa$.}
The following \mcj{result} identifies the mixing distribution. It is a
tempered stable (TS)\mcj{, or power variance}
distribution (\mcj{Tweedie, 1984, Hougaard, 1986, 2000, Fischer \&
Jakob, 2016}). This
distribution has three parameters,  $0 \leq \omega \leq 1$,
$\xi>0$ and $\theta \geq 0$. However, we will take $\theta=1$ throughout
and
refer
to the corresponding distribution as $TS(\omega,\xi)$.
It is defined through its Laplace transform
given by
\[{\cal L}_{\omega,\xi} (s) = \exp\left[\mcj{\xi
\left\{1-(1+s)^\omega \right\}/\omega}\right].\]\smallskip

\noindent {\sc Result 1}.
Let $T|B=b \sim \textrm{PGW}(\mcj{\gamma,\,}\kappa,b)$ and let \(B \sim
\textrm{TS}(\omega,\mcj{\omega\lambda}).\)
Then, $T \sim \textrm{PGW}(\mcj{\gamma,\,\omega \kappa,\,
\lambda})$.\medskip

\noindent {\tt Proof.}
Denote by $g_{\omega,\xi}$ the density of TS$(\omega,\mcj{\omega
\lambda})$.
Then,
\begin{eqnarray*}
P(T \geq t) &=& \int_0^\infty \exp\{-b\,H_N(t^\gamma;\kappa) \}
g_{\omega,\mcj{\omega
\lambda}}(b)
\,db
= {\cal L}_{\omega,\mcj{\omega
\lambda}}\{H_N (t^\gamma;\kappa)\}\\
&=&
\exp\left(\mcj{\lambda} \left[\mcj{1-
(1+t^\gamma)^{\kappa \omega}}\right]\right)=
\exp\left\{-\mcj{\lambda} H_N(t^\gamma; \omega\kappa)\right\}.
\,\,\,\,\,\,\, \Box
\end{eqnarray*}
\smallskip

Some interesting special cases of Result 1 are that:
\begin{itemize}
\item $\omega = 1/2$: if $T|B=b \sim\textrm{PGW}(\mcj{\gamma,\,}\kappa,b)$
and  \(B\)
follows
the inverse Gaussian distribution with parameters $(1/2,1/2)$,
then $T \sim \textrm{PGW}(\mcj{\gamma,\,}\kappa/2, 1)$;
\item $\omega=0$: if $T|B=b \sim\textrm{PGW}(\mcj{\gamma,\,}\kappa,b)$ and
\(B\)
follows
the unit-scale gamma distribution with shape parameter $\xi$,
then $T$ follows the Burr Type XII distribution with power parameter
$\gamma $ and proportionality parameter $\xi \kappa$.
(When $\kappa = 1$, this is the well known result that a  Weibull
distribution with gamma frailty results in the Burr
Type XII distribution.)
\end{itemize}
A related frailty link between the $\kappa=\infty$ and $\kappa=1$
adapted PGW
distributions is:
\begin{itemize}
\item if  $T|B=b$ has c.h.f.\ $b(e^{t^\gamma}-1)$ and
  \(B\) follows
the exponential  distribution with parameter 1, then $T$ follows the
Weibull distribution.
\end{itemize}
\medskip

There is a similar\mcj{, if more complicated,} frailty mixing result for
the APGW distribution but,
because the mixing distribution then depends on $\kappa$, this proves not
to be so convenient for extension to the bivariate case.
\mcj{(It is included in  Appendix A for completeness.)}
Instead, we make
our bivariate extension based on the original PGW distribution and then
change the marginals to APGW distributions, retaining  the
dependence structure associated with the PGW-based extension.
\medskip
\bigskip

\noindent \textbf{3.~The bivariate shared frailty PGW
model}\medskip\medskip

\noindent \textsl{3.1~The model}
\medskip

In the bivariate case, suppose that, for $i=1,2$,
$T_i|B=b \sim {\rm PGW}(\gamma_i,\kappa_i,b)$ \textit{independently} and
$B \sim {\rm TS}(\omega,\omega\lambda)$
with $\lambda>0$ and
$0 \leq \omega \leq 1$. This is a shared frailty model: given the single,
shared, frailty random variable $B=b$, the survival times are
(conditionally) independent; the shared frailty introduces the
dependence, and this dependence is necessarily positive except for the
special case of independence
(Hougaard, 2000, Duchateau \& Janssen, 2008, Wienke, 2011).
In this case, $T_1$ and $T_2$ are (marginally) independent when $\omega =
1$, because then $TS(1,\lambda)$ reduces to a point mass at $\lambda$.
\medskip

The bivariate survival function can be obtained
through an
extension of the proof of \mcj{Result~1.}
Denote by $g_{\omega,\omega \lambda}$ the density of
TS$(\omega,\omega\lambda)$.
Then,
\begin{eqnarray*}
S_N(t_1,t_2 ) &\equiv& P(T_1 \geq t_1,T_2 \geq t_2)\\
 &=& \int_0^\infty P(T_1 \geq t_1;\gamma_1,\kappa_1,b)   \,
P(T_2 \geq t_2;\gamma_2,\kappa_2,b)\,  g_{\omega,\omega
\lambda}(b)\,db\\
&=& \int_0^\infty \exp\left[- b\left\{H_N(t_1;\gamma_1,\kappa_1)+
b\,H_N(t_2;\gamma_2,\kappa_2)\right\}\right] \,
g_{\omega,\omega\lambda}(b)\,db\\
&=& {\cal L}_{\omega,\omega\lambda}\left\{ H_N
(t_1;\gamma_1,\kappa_1)+ H_N
(t_2;\gamma_2,\kappa_2)\right\}\\
&=&
\exp\left(\lambda \left[1- \left\{
(1+t_1^{\gamma_1})^{\kappa_1}+(1+t_2^{\gamma_2})^{\kappa_2}
-1\right\}^\omega\right]\right).
\end{eqnarray*}
The univariate marginals are, of course, $T_1 \sim
{\rm PGW}(\gamma_1,\omega\kappa_1,\lambda)$ and $T_2 \sim\linebreak
{\rm PGW}(\gamma_2,\omega\kappa_2,\lambda)$, by construction.
 This suggests reparametrising by $\tau_i = \omega \kappa_i$, $i=1,2$, in
which case
\begin{equation}S_N(t_1,t_2) =
\exp\left[\lambda
\left\{ 1-L_N^\omega(t_1,t_2)\right\} \right]
\label{SN}\end{equation}
where
\begin{equation}L_N(t_1,t_2)\equiv
(1+t_1^{\gamma_1})^{\tau_1/\omega}
+(1+t_2^{\gamma_2})^{\tau_2/\omega}-1,\label{LN}\end{equation}
so that $T_1 \sim
{\rm PGW}(\gamma_1,\tau_1,\lambda)$, $T_2 \sim
{\rm PGW}(\gamma_2,\tau_2,\lambda)$ and $\omega$ and $\lambda$ control
the
dependence
between \red{$T_1$ and $T_2$}.
This is the bivariate model  with
PGW marginals of interest in this article.\medskip

Just once, we mention the multivariate analogue of a
result, namely
$$ P(T_1 \geq t_1,...,T_p \geq t_p)
=
\exp\left(\lambda \left[1- \left\{\sum_{i=1}^p
(1+t_i^{\gamma_i})^{\tau_i/\omega}
-(p-1)\right\}^\omega\right]\right).
$$ in order to remind the reader that a multivariate version of our
bivariate development is possible, if desired. Indeed, if
$\tau_1=\tau_2=\omega$ in (\ref{SN}), this is a special case
(his $\kappa=1$) of the ``multivariate distribution with Weibull
connections'' of Crowder (1989). Its bivariate form is
$S_{N;C}(t_1,t_2) =\exp\left[\lambda
\left\{ 1-L_{N;C}^\omega(t_1,t_2)\right\} \right]$
where $L_{N;C}(t_1,t_2)\equiv
t_1^{\gamma_1}
+t_2^{\gamma_2}+1.$\medskip\medskip

\noindent \textsl{3.2~Conditional hazard functions}
\medskip

Returning to the bivariate case, all
joint and conditional density, survival  and hazard functions are
available in closed form. We will explicitly look at two conditional
distributions. For ease of notation, write
 $K(t_2) = (1+t_2^{\gamma_2})^{\tau_2}.$
Then, it is immediate from (\ref{SN}) that a first conditional survival
function
is
$$P(T_1\geq t_1|T_2 \geq t_2) =  e^{\lambda
K(t_2)}\,\exp\left\{ -\lambda L_N^\omega(t_1,t_2)\right\}$$
and the associated conditional c.h.f.\ is
$H(t_1|T_2 \geq t_2)=\lambda\left\{L_N^\omega(t_1,t_2) -K(t_2)  \right\}.$
Writing $$L_N^{10}(t_1,t_2) = \partial L_N(t_1,t_2)/\partial t_1 =
\gamma_1\tau_1
t_1^{\gamma_1-1}(1+t_1^{\gamma_1})^{(\tau_1/\omega)-1}/\omega,$$
the conditional hazard function
\begin{equation}h(t_1|T_2 \geq t_2) = \lambda \omega
L_N^{\omega-1}(t_1,t_2)\,L_N^{10}(t_1,t_2)\label{hazge}\end{equation}
is seen to behave as
$t_1^{\gamma_1-1}$ as $t_1 \to 0$ and as $ t_1^{\gamma_1 \tau_1-1}$ as
$t_1 \to \infty$, a property shared with the
marginal hazard function to which $h(t_1|T_2 \geq t_2)$ corresponds when
$t_2=0$. Now,
\begin{equation}h'(t_1|T_2 \geq t_2) = \lambda \omega
L_N^{\omega-2}(t_1,t_2)
\left[
(\omega-1)\{L_N^{10}(t_1,t_2)\}^2+L_N(t_1,t_2)\,L_N^{20}(t_1,t_2)\right]
\label{hazgederiv}\end{equation}
where
\begin{eqnarray*}
L_N^{20}(t_1,t_2) &=&
\partial^2L_N(t_1,t_2)/\partial
t_1\partial t_2
\\&=& {\gamma_1\tau_1}
t_1^{\gamma_1-2}(1+t_1^{\gamma_1})^{(\tau_1/\omega)-2}
\left\{ \gamma_1-1+\left( \frac{\gamma_1\tau_1}\omega
-1\right)t_1^{\gamma_1}\right\}/{\omega} .\end{eqnarray*}
It then follows that $h'(t_1|T_2 \geq t_2)$ is equal to positive terms
times
\begin{eqnarray*}
&&
\left\{(1+t_2^{\gamma_2})^{\tau_2/\omega}-1\right\} \left\{ \gamma_1-1
+\left( \frac{\gamma_1\tau_1}{\omega}-1\right) t_1^{\gamma_1}\right\} \\
&&
~~~~~~~~~~~~~~~~~~+
(1+t_1^{\gamma_1})^{\tau_1/\omega} \left\{ \gamma_1-1
+\left({\gamma_1\tau_1}-1\right)t_1^{\gamma_1}\right\}.
\end{eqnarray*}
When $t_2>0$, all we can guarantee is that
$${\rm if}~~\gamma_1 \geq 1~~{\rm and}~~ \gamma_1\tau_1 \geq 1~~{\rm
then} ~~h(t_1|T_2 \geq t_2) ~~\textrm{is increasing},$$
corresponding to the $t_2=0$ case, and
$${\rm if}~~\gamma_1 \leq 1~~{\rm and}~~ \gamma_1\tau_1 \leq \omega~~{\rm
then} ~~h(t_1|T_2 \geq t_2) ~~\textrm{is decreasing},$$
a more restricted parameter range than in the marginal case. Otherwise, we
cannot guarantee no more complicated shapes than in the $t_2=0$ case even
though the hazards start and end up with the same behaviour whatever the
value of $t_2$.
\medskip

Another conditional survival function is
\begin{eqnarray*}
P(T_1\geq t_1|T_2 = t_2) &=& K^{(1/\omega)-1}(t_2) \,e^{\lambda
K(t_2)}
\, L_N^{\omega-1}(t_1,t_2)\,
   \exp\left\{-\lambda  L_N^\omega(t_1,t_2)\right\}\\
&=& K^{(1/\omega)-1}(t_2)
\, L_N^{\omega-1}(t_1,t_2)\, P(T_1\geq t_1|T_2 \geq t_2).
\end{eqnarray*}
It follows that
\begin{equation}h(t_1|T_2=t_2) = h(t_1|T_2 \geq t_2) +(1-\omega)
L_N^{10}(t_1,t_2)/L_N(t_1,t_2),\label{hazeq}\end{equation}
which has the same limiting behaviour as $h(t_1|T_2 \geq t_2)$ given by
 (\ref{hazge}).
Also,
\begin{eqnarray}
h'(t_1|T_2=t_2)&=&h'(t_1|T_2 \geq t_2) +(1-\omega) \{L_N(t_1,t_2)\}^{-2}
\nonumber \\
&&~~~~~~~\times
\left[L_N(t_1,t_2)\,L_N^{20}(t_1,t_2) -\{L_N^{10}(t_1,t_2)\}^2\right].
\label{hazeqderiv}
\end{eqnarray}
The term in square brackets is  equal to positive terms times
$$\left\{(1+t_2^{\gamma_2})^{\tau_2/\omega}-1\right\} \left\{ \gamma_1-1
+\left( \frac{\gamma_1\tau_1}{\omega}-1\right) t_1^{\gamma_1}\right\}+
(1+t_1^{\gamma_1})^{\tau_1/\omega} \left(
\gamma_1-1-t_1^{\gamma_1}\right) .$$
When $t_2>0$, we can only be sure that,
just as for
the other conditional hazard function,
$${\rm if}~~\gamma_1 \leq 1~~{\rm and}~~ \gamma_1\tau_1 \leq \omega~~{\rm
then} ~~h(t_1|T_2 = t_2) ~~\textrm{is decreasing}.$$
There is, however, no  simple guarantee of increasingness.
\medskip\medskip

\noindent \textsl{3.3~Clayton's cross ratio dependence function}
\medskip

Arguably the most important measure of pointwise dependence
for
bivariate survival distributions is the Clayton (1978) cross ratio
dependence function, one of
whose formulations is
$$\theta(t_1,t_2) = \frac{h(t_1|T_2=t_2)}{h(t_1|T_2\geq t_2)}$$
(see also Oakes, 1989, Hougaard, 2000, Duchateau \& Janssen, 2008).
Recall that, loosely speaking, positive (pointwise) dependence
corresponds to values of $\theta(t_1,t_2)>1$.
Then, from  formulae in Section 3.2,
\begin{equation}\theta_N(t_1,t_2) =
1+\frac{1-\omega}{\lambda\omega}\,\frac1{\{L_N(t_1,t_2)\}^\omega}
 =
1+\frac{1-\omega}{\omega}\,\frac1{\lambda-\log
S_N(t_1,t_2)}.\label{clayton}\end{equation}
For each $t_1$ and $t_2$, as $\lambda$ increases,
$\theta_N(t_1,t_2)$ decreases through values greater than 1, with
$\lim_{\lambda\to\infty}\theta_N(t_1,t_2)
= 1$; the latter
corresponds to an uninteresting degenerate independence case.
Clearly, in the non-degenerate independence case when $\omega = 1$ (with proper
PGW marginals),  we have
 $\theta_N(t_1,t_2)=1$.
It can also be proved that $\theta_N(t_1,t_2)$ decreases with $\omega \in
(0,1)$, again through values greater than 1. To this end,
write $r_i = (1+t_i^{\gamma_i})^{\tau_i}$, $i=1,2$, so that $L_N(t_1,t_2) =
r_1^{1/\omega}+r_2^{1/\omega}-1$.
Then we need the
derivative with respect to $\omega$ of $(1-\omega)/[\omega
\{L_N(t_1,t_2)\}^\omega]$ which is a
positive quantity times
$$-1-\omega(1-\omega) \left\{ \log
(r_1^{1/\omega}+r_2^{1/\omega}-1) -
\frac{(r_1^{1/\omega}\log
r_1+r_2^{1/\omega}
\log r_2)}{\omega(r_1^{1/\omega}+r_2^{1/\omega}-1)}
\right\}.$$
The term in curly brackets can be seen to be
positive by consideration of the function
$(a_1+a_2-1)\log(a_1+a_2-1)-a_1\log a_1-a_2\log a_2$ to which the term is
proportional when $a_i= r_i^{1/\omega}$, $i=1,2$.
But $\min\{a_1,a_2\}>1$ and, by differentiation, the function is
increasing in
$a_1$ and $a_2$. It therefore tends to its infimum value  when
$a=b=1$, and this infimum value is zero.
The derivative is therefore negative, and the Clayton cross ratio
dependence function decreases in $\omega$ for all $t_1,t_2$, as suggested.
\medskip\medskip

\noindent \textsl{3.4~The survival copula}
\medskip

Dependence in the bivariate PGW distribution can also be understood
through its
survival copula. This is the cumulative distribution function (c.d.f.) defined by
$$ \widehat{C}(u,v) = S_N(S_{1,N}^{-1}(u),S_{2,N}^{-1}(v))
, ~~~0<u,v<1,$$
with uniformly distributed marginals; here, $S_{i,N}$ is the $i$th
marginal
survival function, $i=1,2.$ It is easily seen that in this case
\begin{equation}
\widehat{C}(u,v) = \exp\left[\lambda-\left\{(\lambda-\log
u)^{1/\omega}+
(\lambda-\log
v)^{1/\omega}-\lambda^{1/\omega}\right\}^\omega\right].\label{copula}
\end{equation}
(Survival function (\ref{SN}) is reconstructed from (\ref{copula}) as
$\widehat{C}(S_{1,N}(t_1),S_{2,N}(t_2))$.)\medskip

This is a well known, Archimedean, copula. It is
 the  BB9
copula of Joe (1997, 2015), also known as the PVF (power variance
function) copula (Hougaard, 2000, Duchateau \& Janssen, 2008,  Romeo,
Meyer \& Gallardo, 2018). Of course,
$\omega=1$ corresponds to the case of independence $(\widehat{C}(u,v) =
uv$) while as $\omega \to 0$, $\widehat{C}(u,v) \to \min(u,v)$, the
Fr\'echet
upper bound (equivalent to $T_1=T_2$). We also have
independence as $\lambda \to \infty$\mage{. When $\lambda =1$, the BB9/PVF copula
reduces to copula (4.2.13) of Nelsen (2006)
 while,} as $\lambda \to 0$, \mage{it}
tends to the Gumbel copula
$$\widehat{C}_G(u,v) = \exp\left[-\left\{(-\log u)^{1/\omega}+(-\log
v)^{1/\omega}\right\}^\omega \right].$$

\textit{The novelty in this article is that we  \textit{naturally} use the
BB9/PVF
copula
| as
opposed to any other arbitrarily chosen  copula | in
conjunction with PGW marginal distributions | as
opposed to any other arbitrarily chosen marginals |   because
univariate PGW distributions are closed under
tempered stable/power variance frailty
distributions.}\medskip

Concordance, or positive quadrat dependence, is a strong
positive dependence property that implies many others (for instance, if
concordance increases, so do Kendall's tau, Spearman's rho, and tail
dependence). Joe (2014) shows that, for the BB9/PVF copula,
concordance increases as either $\omega$ or $\lambda$
decreases. We will briefly investigate some more specifics of the
behaviour of
Kendall's tau, $K$ say, and Spearman's rho, $S$ say, as functions of
$\omega$ and $\lambda$.
\medskip

\begin{figure}
\begin{center}
\includegraphics[width=9.5cm,height=8cm]{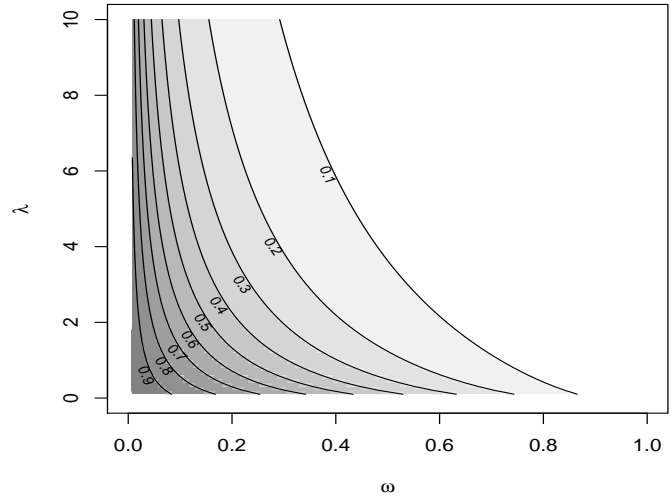}
\caption{Kendall's tau plotted as a function of $0 \leq \omega \leq 1$
and $\lambda>0$ (curtailed at $\lambda=10$) for the BB9/PVF copula.}
\label{Kendall-tau}
\end{center}
\end{figure}

A bivariate  Archimedean copula is of the form
$\phi\{\phi^{-1}(u)+\phi^{-1}(v)\}$ where $\phi$ is a survival function on
$\mathbb{R}^+$ associated
with a decreasing density. For BB9/PVF,
$ \phi(s) = \exp\left[ \lambda \left\{1- (1+s)^\omega \right\}\right]$
and
\begin{eqnarray*}
K& =& 1-4 \int_0^\infty s \{\phi'(s)\}^2ds\\
&=& 1- \omega \left\{1+2\lambda-(2\lambda)^{1/\omega} e^{2\lambda} \,
\Gamma(2-(1/\omega),2\lambda)\right\}\nonumber
\end{eqnarray*}
where $\Gamma(a,z) = \int_z^\infty x^{a-1}e^{-x}\,dx$ is an
incomplete
gamma function (for alternative versions of this formula, see
(7.58) of Hougaard
(2000) and (4.87) of Duchateau \& Janssen, 2008).
$K$ decreases from 1 (Fr\'echet copula)
towards 0 (independence) as $\omega$
increases from 0 to 1 and
from $1-\omega$ (Gumbel copula)
towards 0 (independence) as $\lambda$
increases from 0  without bound. (It is easy to confirm that $K \leq
1-\omega$
for all $\lambda>0,0\leq \omega \leq 1$ using a standard inequality
for
the incomplete gamma function which is a simple consequence of
integration by parts.)
The behaviour of $K$ is confirmed in Figure~\ref{Kendall-tau}  which is
a
contour plot of $K$ as a function of $\omega$ (horizontal axis)
and $\lambda$ (vertical axis).

\begin{figure}
\begin{center}
\includegraphics[width=9.5cm,height=8cm]{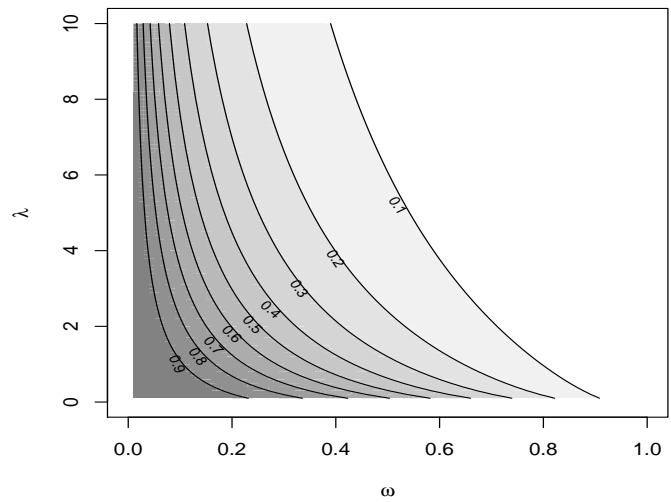}
\caption{Spearman's rho plotted as a function of $0 \leq \omega \leq 1$
and $\lambda>0$ (curtailed at $\lambda=10$) for the BB9/PVF copula.}
\label{Spearman-rho}
\end{center}
\end{figure}

\medskip
There is no such similar explicit form
for  Spearman's rho for Archimedean copulas in general or the BB9/PVF
copula in particular,  but a numerically-derived
graphical representation of it for BB9/PVF as a
function of
$\omega$ and $\lambda$ is also of interest; see
Figure~\ref{Spearman-rho}. The qualitative features of
Figure~\ref{Spearman-rho} are very similar to those of
Figure~\ref{Kendall-tau}: $S$ too decreases from 1 (Fr\'echet copula)
towards 0 (independence) as $\omega$
increases from 0 to 1 and
from a positive value (Gumbel copula)
towards 0 (independence) as $\lambda$
increases from 0  without bound.  It is also suggested that $S \geq K$ for
the BB9/PVF copula, but we have no proof.
An upper bound for $S$ for the Gumbel copula, and hence for the BB9/PVF copula,
is \begin{equation}
S \leq
\min\left[3\left\{\frac{2^{2(2-\omega)}}{(1+2^{1-\omega})^2}-1\right\},1\right].\label{bound}\end{equation}
The bound is non-trivial for $\omega>1+\log_2(\sqrt{3}-1) \simeq 0.55.$
See Appendix \mcj{B} for proof of this.


\medskip
\bigskip

\noindent \textbf{4.~The bivariate APGW
model}\medskip\medskip

\noindent \textsl{4.1~The model}
\medskip

We cannot simply replace the steps taken for the original PGW
distribution for the adapted PGW distribution despite the existence of a
parallel
frailty result for APGW distributions in
Result~A1 in \mcj{Appendix A}: in Result~A1, the
TS distribution depends on $\kappa$, so the frailty distribution involved
is not a single one to be shared by each failure time.
\medskip

There is, however, a (still tractable) alternative which is to
transform each marginal random variable, $T_i$, currently following an
ordinary
${\rm PGW}(\gamma_i,\tau_i,\lambda)$ distribution to one, $Z_i$ say,
following an ${\rm APGW}(\gamma_i,\tau_i,\lambda)$ distribution, using
$$ Z_i = \left[(\tau+1)^{1-(1/\tau)}\left\{1+{\tau}
(1+t^\gamma)^\tau\right\}^{1/\tau}-(\tau+1)\right]^{1/\gamma},$$
$i=1,2.$ The resulting bivariate APGW distribution has
survival function \begin{equation}S_A(z_1,z_2) =
\exp\left[\lambda
\left\{ 1-L_A^\omega(z_1,z_2)\right\} \right]
\label{SA}\end{equation}
where
\begin{eqnarray}
L_A(z_1,z_2)&=&
\left\{\frac{(\tau_1+1)}{\tau_1}
\left(1+\frac{z_1^{\gamma_1}}{\tau_1+1}\right)^{\tau_1}-\frac1{\tau_1}
\right\}^{1/\omega} \nonumber\\
&&~~~~~~~~+
\left\{\frac{(\tau_2+1)}{\tau_2}
\left(1+\frac{z_2^{\gamma_2}}{\tau_2+1}\right)^{\tau_2}-\frac1{\tau_2}\right\}^{1/\omega}
-1.\label{LA}
\end{eqnarray}
Of course, this is nothing other than introducing APGW marginals instead
of PGW marginals into the BB9/PVF copula (\ref{copula}). (Survival
function (\ref{SA}) is
$\widehat{C}(S_{1,A}(z_1),S_{2,A}(z_2))$
where $S_{i,A}$ is the $i$th
marginal APGW
survival function, $i=1,2.$)\medskip

Special and limiting cases include bivariate Weibull
($(\tau_1,\tau_2)=(1,1)$), log-logistic/Burr Type
XII ($(\tau_1,\tau_2) \to (0,0)$) and Gompertz
distributions ($(\tau_1,\tau_2)$\linebreak $\to
(\infty,\infty)$). Their survival functions are all of the form (\ref{SA})
with $L_A$ functions given by
$$L_{A;W} (z_1,z_2)=
(1+z_1^{\gamma_1})^{1/\omega}+(1+z_2^{\gamma_2})^{1/\omega}-1,$$
$$L_{A;B}(z_1,z_2)=
\left\{1+\log(1+z_1^{\gamma_1})\right\}^{1/\omega}+
\left\{1+\log(1+z_2^{\gamma_2})\right\}^{1/\omega}
-1$$
and
$$L_{A;G}(z_1,z_2)= \exp(z_1^{\gamma_1}/\omega)+
\exp(z_2^{\gamma_2}/\omega)-1,$$
respectively.
\medskip

Because the bivariate APGW distribution of this section has the same
copula as the
bivariate  PGW
distribution of Section 2, all those aspects of dependence that are
encapsulated in the
copula, and explored in Section 3.4, apply to the bivariate APGW
distribution in the
same way as they do to the bivariate PGW distribution.
\medskip\medskip

\noindent \textsl{4.2~Conditional hazard functions and Clayton's  cross
ratio dependence function}\medskip

The investigations of Sections 3.2 and 3.3, which are not purely
copula-\linebreak dependent, need to be reworked for the
bivariate APGW
distribution. Because survival function $(\ref{SA})$ has the same form as
survival function (\ref{SN}), the conditional hazard functions and
Clayton's cross ratio dependence function have the same form as before
(\ref{hazge}), (\ref{hazeq}) and (\ref{clayton}), respectively, but
with $L_A$ given by (\ref{LA}) replacing $L_N$ given by (\ref{LN}).
\medskip

The claims made for the bivariate APGW distribution in the remainder of
this section are based on formulae
given in Appendix~\mcj{C}. We find that for $h(z_1|Z_2 \geq z_2)$ we have:
$${\rm if}~~\gamma_1 \geq 1~~{\rm and}~~ \gamma_1\tau_1 \geq 1~~{\rm
then} ~~h(z_1|Z_2 \geq Z_2) ~~\textrm{is increasing},$$
corresponding to the $z_2=0$ case,
and
$${\rm if}~~\gamma_1 \leq 1~~{\rm and}~~ \gamma_1 \leq
\frac{\omega}{1-\omega+\tau_1}~~{\rm
then} ~~h(z_1|Z_2 \geq z_2) ~~\textrm{is decreasing},$$
which is a little more restricted than in the bivariate PGW case.
And also, in the APGW case, we can
only be sure that
$${\rm if}~~\gamma_1 \leq 1~~{\rm and}~~ \gamma_1 \leq
\frac{\omega}{1-\omega+\tau_1}~~{\rm
then} ~~h(x|Y = y) ~~\textrm{is decreasing}$$
but, as in the bivariate PGW case, there is  no simple  guarantee of
positivity.
\medskip

As trailed above, the cross ratio dependence function is now
$$\theta_A(z_1,z_2) =
1+\frac1\lambda\,\left(\frac1\omega-1\right)\,\frac1{\{L_A(z_1,z_2)\}^\omega}.$$
Its monotonicity properties --- decreasingness in $\lambda$ and $\omega$
--- remain as in the PGW case. The latter now follows by the same argument
as in Section 3.3 except with  $r_i =
[(\tau_i+1)\{1+z_i^{\gamma_1}/(\tau_i+1)\}^{\tau_i}-1]/\tau_i$, $i=1,2.$
\medskip
\bigskip

\noindent \textbf{5.~Application}\medskip\medskip

The well-known retinopathy dataset of Huster, Brookmeyer, and Self
(1989; henceforth HBS) (available in the \texttt{survival} package in
\ddd{\textsf{R}}) consists
of measurements of the time to blindness in each of their eyes for 197
individuals. For each individual,
one eye was randomised to a laser treatment with the other eye acting as
a
control and, hence, survival times are naturally paired within
individuals. The main focus of this study was to evaluate the
effectiveness of the treatment, that is, to compare aspects of the
marginal survival distributions (in the presence of dependence). However,
 a covariate indicating the diabetes
type (juvenile or adult at age of onset) was also of interest.
\medskip

We use the
bivariate APGW distribution to analyse these data, first without the
covariate (Section 5.1) and then with the addition of the covariate
(Section 5.2). The `full' bivariate APGW distribution that we consider is
that of Section 4 with horizontal
scale/AFT
parameters added into each marginal, that is,
the model with survival function $S_A(\phi_1 z_1,\phi_2z_2)$ where $S_A$
is given by (\ref{SA}) and (\ref{LA}). This model has eight parameters:
\begin{align}
 (Z_1, Z_2) \sim
\text{bivariate~APGW}(\underbrace{\lambda,\omega,}_{\text{dependence}}
~\underbrace{\phi_1,\gamma_1,\tau_1}_{Z_1~\text{marginal}},
~\underbrace{\phi_2,\gamma_2,\tau_2}_{Z_2~\text{marginal}}~).
\label{fullmodel}\end{align}
However, for ease of optimisation in fitting models using maximum
likelihood estimation, we will prefer to work in terms
of the vector of unconstrained parameters, $\theta = (\theta_\lambda,
\theta_\omega,
\theta_{\phi_1}, \theta_{\gamma_1}, \theta_{\tau_1},  \theta_{\phi_2},
\theta_{\gamma_2}, \theta_{\tau_2})^T$ where
\begin{eqnarray*}
\theta_\lambda &=& \log \lambda,\qquad
\theta_\omega = \log\{\omega/(1-\omega)\},\qquad
\theta_{\phi_j} = \log \phi_j,\\
\theta_{\gamma_j} &=& \log \gamma_j,\qquad\quad
\theta_{\tau_j} = \log(\tau_j+1),
\end{eqnarray*}
\ddd{$j=1,2$. In the context of the retinopathy dataset, the index `1'
corresponds to treatment and the index `2' to control.}

\medskip\smallskip
\noindent \textsl{5.1~Treatment effect (no covariate)}
\medskip

In this subsection, we consider a series of submodels of
(\ref{fullmodel}) investigating the
treatment effect (without including the diabetes covariate). Table
\ref{tab:trtmods} compares the eight models arising from all combinations
of the following constraints:
\begin{enumerate}
\item[(i)] common scale
$\phi_1 = \phi_2 = \phi$,
\item[(ii)] common power $\gamma_1 = \gamma_2 =
\gamma$, and
\item[(iii)] common distribution $\tau_1 = \tau_2 = \tau$,
\end{enumerate}
together with, as the first model, the unconstrained model. Each model has
been fitted to the data using maximum likelihood\ddd{, implemented using
a Newton procedure
via the standard {\tt nlm} optimiser in \textsf{R}.}

\medskip

\begin{table}[htbp]
\begin{center}
\caption{Comparison of treatment models\label{tab:trtmods}}
\vspace{0.05cm}
\begin{tabular}{ccccrrrrr}
\hline
Model &  Common & dim$(\theta)$ &
$\ell(\theta)$ &  AIC & BIC & $\Delta_{\text{AIC}}$ &
$\Delta_{\text{BIC}}$ & $K$ \\
\hline
1     &    ---               &     8 & -824.24  & 1664.48  & 1690.74  &  2.36  &  8.93 & 0.24 \\
2     &  $\phi$              &     7 & -824.62  & 1663.25  & 1686.23  &  1.13  &  4.41 & 0.28 \\
3     &  $\gamma$            &     7 & -824.64  & 1663.28  & 1686.27  &  1.17  &  4.45 & 0.24 \\
4     &  $\tau$              &     7 & -824.71  & 1663.42  & 1686.40  &  1.30  &  4.59 & 0.18 \\
5     &  $\phi,\gamma$       &     6 & -826.45  & 1664.91  & 1684.61  &  2.79  &  2.79 & 0.18 \\
6     &  $\phi,\tau$         &     6 & -826.01  & 1664.02  & 1683.72  &  1.90  &  1.90 & 0.19 \\
7     &  $\gamma,\tau$       &     6 & -825.06  & 1662.12  & 1681.81  &  0.00  &  0.00 & 0.18 \\
8     &  $\phi,\tau,\gamma$  &     5 & -839.59  & 1689.19  & 1705.60  & 27.07  & 23.79 & 0.16 \\
\hline
\end{tabular}
\footnotesize{``Common'' indicates which parameters (``pars'') are
constrained to
 be equal,\\ $\ell(\theta)$ is the log-likelihood value,
 $\Delta_{\text{AIC}} =
 \text{AIC} - \min(\text{AIC})$, $\Delta_{\text{BIC}} = \text{BIC} -
 \min(\text{BIC})$,\\ $K$ is Kendall's tau.}
\end{center}
\end{table}

\begin{figure}[htbp]
\begin{center}
\includegraphics[width=0.8\textwidth, trim = 0.0cm 0.4cm 0.5cm 0.8cm,
clip]{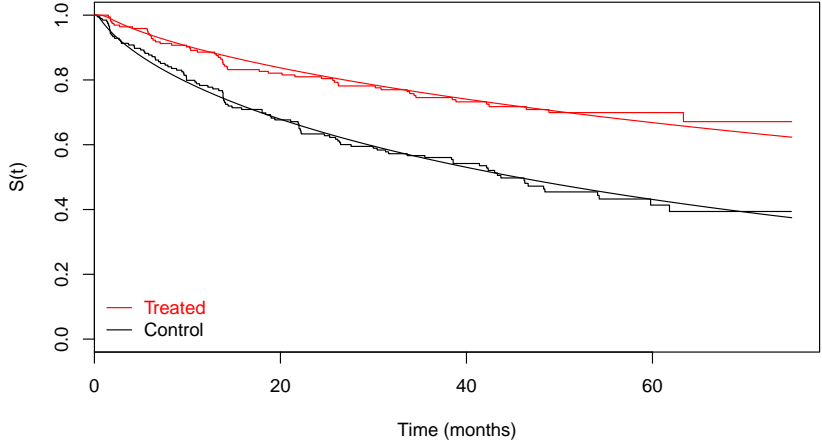}
\caption{Kaplan--Meier (step) curves with model-based curves (smooth)
overlaid for Model 7 of Table \ref{tab:trtmods}.}\label{fig:KMfit}
\end{center}
\end{figure}

We see that both the AIC and BIC are considerably larger for Model 8
than for any of the other fitted models. Model 8
is the model with equal distributions for the treated and untreated
groups; hence, there certainly appears to be a difference between these
groups.
On the other hand, the full flexibility of Model 1 with unconstrained
parameters
 appears not to be required here.  The model with both the lowest AIC and
BIC is Model 7 which has common shape parameters,  but different scales,
that is, the
 basic  distribution and shape of hazard is the same for these two
treatment groups,
but they have different scales; Figure \ref{fig:KMfit}  reveals that the
fit to the
data is excellent.\medskip

Table \ref{tab:phimod} displays parameter estimates for Model 7. As the
two treatment groups here differ only with respect to their horizontal
scale, we have the AFT property across groups. Thus, the quantile ratio is
given by $\psi = \exp(\theta_{\phi_2}-\theta_{\phi_1})$ which is estimated
as $\hat\psi = 2.84$, that is, time to blindness is almost 3 times
longer in eyes that receive laser treatment than in those that do not;
the corresponding 95\% confidence interval is $(1.64,4.92)$ (calculated
using the delta method).\medskip

\begin{table}[htbp]
\begin{center}
\caption{Parameter estimates using Model 7\label{tab:phimod}}
\vspace{0.05cm}
\begin{tabular}{lcccccc}
\hline
Parameter & $\theta_\lambda$ &$\theta_\omega$ &$\theta_\gamma$ &
$\theta_\tau$ & $\theta_{\phi_1}$ &
$\theta_{\phi_2}$ \\
\hline
Estimate & $-5.57$ & 1.52 & 1.52 & 0.14 & $-0.07$ & 0.98\\
(S.E.) & (2.02) & (0.40) & (1.78) & (0.23) & (0.45) & (0.62)\\
\hline
\end{tabular}

\footnotesize{Parameters shown are in unconstrained form, for example,
$\theta_\lambda = \log\lambda$.\\$\phi_1$
and $\phi_2$  are the scale
parameters for the treatment and control group,
respectively.}
\end{center}
\end{table}

Note that all models \mage{fitted in Table~\ref{tab:trtmods}} indicate \mage{quite a
low} level
of correlation as measured by Kendall's tau.
\kevin{In practice,} \mage{however,} \kevin{it is often the case that the marginal
models are fixed,}
\mage{for example, to
Weibull or Gompertz distributions}\kevin{. In contrast, our more flexible} \mage{APGW
marginal distribution} \kevin{automatically selects the
marginal models via the $\tau$}
\mage{parameter(s);}  the
estimated (common) $\tau$ value in our
model is 0.15, with 95\%
confidence interval $(-0.26,0.78)$, which does not support Weibull \mage{or Gompertz}
margins
($\tau=1$\mage{, $\tau=\infty$}). \mage{Misspecified} \kevin{marginal models
have
the potential to bias the measure of dependence.
We briefly investigate the effect of the marginal model on Kendall's tau by starting
with Model 7 (of Table \ref{tab:phimod}) and profiling over the $\tau$ parameter
(including key models such as log-logistic, Weibull and Gompertz). The results are
shown in Table \ref{tab:proftau} where we see that it is indeed the case that the
value of $K$ varies with the choice of marginal model. In particular, the
maximum} \mage{likelihood} \kevin{marginal model with $\hat\tau = 0.15$ yields a
lower $K$ value}
\mage{than}
\kevin{that
of other $\tau$ values. Hence, fixing $\tau$} \mage{a priori} \kevin{has the
potential to alter our
view of the level of dependence.
Interestingly, a 95\% confidence interval for
$K$ based on Model 7 is} \mage{$(0.08,0.31)$}\kevin{. While this} \mage{covers}
\kevin{all $K$
values in Table
\ref{tab:proftau}} \mage{(just!), note from} \kevin{the Chi-squared statistics
of Table
\ref{tab:proftau} that higher $K$ values in models with Weibull or Gompertz margins
are not supported by the data.}

\begin{table}[htbp]
\kevin{\begin{center}
\caption{Profiling $\tau$ for Model 7\label{tab:proftau}}
\vspace{0.1cm}
\begin{tabular}{ccccc}
\hline
Model & $\tau$ & $K$ & $\ell(\theta)$ & $\chi^2_1$ \\
\hline
Log-logistic & 0.00 & 0.27 & -826.89 & 3.66 \\
             & 0.07 & 0.18 & -825.24 & 0.36 \\
     Model 7 & 0.15 & 0.18 & -825.06 & 0.00 \\
             & 0.57 & 0.20 & -826.74 & 3.36 \\
     Weibull & 1.00 & 0.24 & -829.03 & 7.94 \\
             & 1.23 & 0.27 & -829.45 & 8.78 \\
             & 1.72 & 0.31 & -829.60 & 9.09 \\
Gompertz & $\infty$ & 0.31 & -829.63 & 9.14 \\
\hline
\end{tabular}

\footnotesize{$\chi^2_1$ is the chi-squared statistic given by computing
$2(\ell_{\text{Model 7}} - \ell_{\tau_0})$ where $\ell_{\text{Model 7}}$\\ is the
likelihood
for Model 7, and $\ell_{\tau_0}$ is the likelihood for $\tau$ fixed at $\tau_0$.}
\end{center}} \end{table}

Note that the best-fitting treatment model of
HBS is also one in which treatment enters
the scale but not the shape. Their model comprises Weibull marginals
together with a Clayton copula. It too has an AFT interpretation,
their estimated quantile ratio turning out to be
$2.62$, which is numerically similar to our
result.
 The AIC and BIC values for HBS's model
are, respectively, 1667.21 and 1680.34, that is, the AIC value is
much higher
than that of
our
Model 7, while the BIC is slightly lower. The value of $K$
associated with HBS's
best-fitting model is 0.30, which, in light of Table 3, may be
somewhat high. This could be a result of
fixing to Weibull marginals or the use of a different (one-parameter)
copula. Thus, while our proposed model can adapt readily to a variety of
situations through its flexible copula and margins, the general use of
simpler copula and marginal components will not work well in as many
cases.

\medskip\bigskip
\noindent \textsl{5.2~Diabetes effect (added covariate)}
\medskip

  From the previous subsection, there is clearly a difference between
treatment groups which manifests via a scale change rather than a shape change. It is also
 of interest to discover whether or not the type of diabetes --
``juvenile'' (the reference group here) or ``adult'' -- is related to
survival and, indeed,
 whether or not the diabetes effect interacts with the treatment effect.
We  will investigate this by extending the best treatment model from the previous
 section, Model 7, as follows:
$$
\theta_\gamma = \theta_{\gamma,0} + \theta_{\gamma,1} D,\qquad
\theta_{\phi_1} = \theta_{\phi_1,0} + \theta_{\phi_1,1} D,\qquad
\theta_{\phi_2} = \theta_{\phi_2,0} + \theta_{\phi_2,1} D,
$$
  where $D$ is the binary diabetes indicator such that $D=1$ means that
 the diabetes type  is adult. Note that, in line with our
previous work,
 we are
  keeping the distributional parameter, $\tau$, as a covariate-independent
 parameter. Moreover, the copula parameters will  also  remain independent
of the covariate \mage{for the moment}.\medskip

\begin{figure}[htbp]
\begin{center}
\includegraphics[width=0.8\textwidth, trim = 0.0cm 0.4cm 0.5cm 0.8cm,
clip]{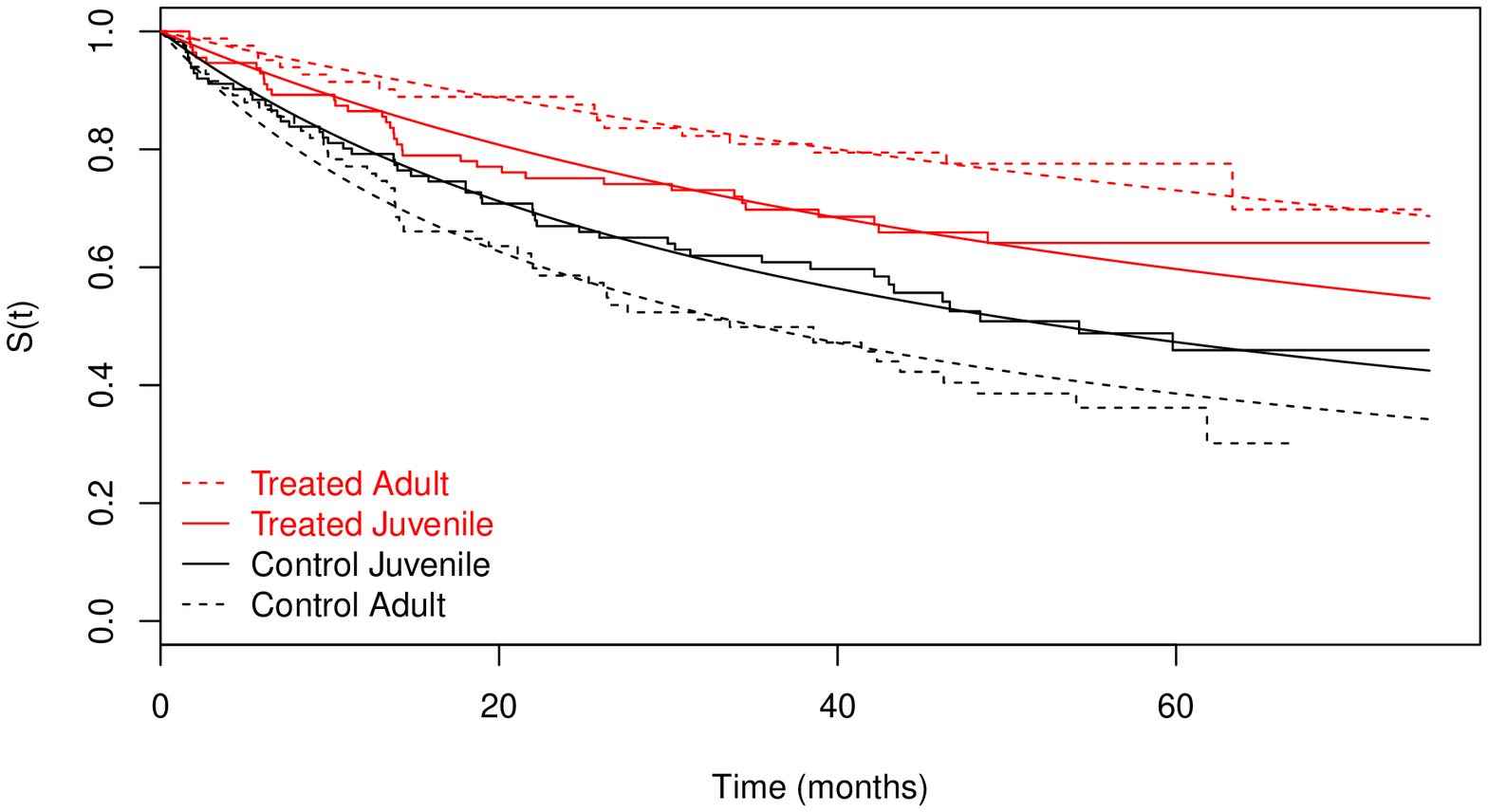}
\caption{Kaplan--Meier  (step) curves with model-based curves (smooth) overlayed for
Model 7(b) of Table \ref{tab:trtmodsD}}\label{fig:KMfitD}
\end{center}
\end{figure}

 The above model set-up permits a diabetes effect which interacts with
treatment and, indeed, non-AFT effects via the inclusion of $D$ into the
power shape parameter. We  arrive at four models of interest characterized
by combinations of:
 \begin{enumerate}
 \item[(i)] $D$ has a non-AFT effect ($\theta_{\gamma,1}$ is free) or $D$
has
 an AFT effect ($\theta_{\gamma,1} = 0$);
 \item[(ii)] $D$ interacts with treatment ($\theta_{\phi_1,1}$ and
  \red{$\theta_{\phi_2,1}$} are free) or not
($\theta_{\phi_1,1}=\,\red{\theta_{\phi_2,1}}\,=0$).
\end{enumerate}
  These four models were fitted to the data using maximum likelihood, and
the results are summarised in Table
\ref{tab:trtmodsD}. Model 7(b) has the lowest AIC and BIC values. Hence, it
  appears that diabetes interacts with treatment, and can be described by
an AFT effect, that is, diabetes does not affect the shape of the
distribution. This is
  in line with the diabetes model considered by HBS --- although they did
not investigate the potential shape effect of diabetes.  We can see
from Figure
 \ref{fig:KMfitD} that
Model 7(b) provides an excellent fit to the data.
\medskip

\begin{table}[htbp]
\begin{center}
\caption{Comparison of treatment models with diabetes covariate\label{tab:trtmodsD}}
\vspace{0.1cm}
\begin{tabular}{ccccrrrrr}
\hline
Model &  Diabetes Effect & $\dim(\theta)$ &
$\ell(\theta)$ &  AIC & BIC & $\Delta_{\text{AIC}}$ & $\Delta_{\text{BIC}}$ & $K$ \\
\hline
7(a)  &   non-AFT int. &     9 & -820.86  & 1659.73  & 1689.28  &  0.38  &
3.67 & 0.19 \\
7(b)  &   AFT int.     &     8 & -821.67  & 1659.35  & 1685.61  &  0.00  &
0.00 & 0.19 \\
7(c)  &   non-AFT             &     8 & -822.74  & 1661.47  & 1687.74  &  2.13  &  2.13 & 0.17 \\
7(d)  &   AFT                 &     7 & -825.04  & 1664.08  & 1687.06  &  4.73  &  1.45 & 0.18 \\
\hline
\end{tabular}

 \footnotesize{``int.'' is short for ``interaction'', $\ell(\theta)$ is
the log-likelihood value,
$\Delta_{\text{AIC}} = \text{AIC} - \min(\text{AIC})$, $\Delta_{\text{BIC}} = \text{BIC} -
 \min(\text{BIC})$, $K$ is Kendall's tau.}
\end{center}
\end{table}

\begin{table}[htbp]
\begin{center}
\caption{Parameter estimates using Model 7(b)\label{tab:phimodD}}
\vspace{0.05cm}
\begin{tabular}{lcccccccc}
\hline
Parameter & $\theta_\lambda$ &$\theta_\omega$ &$\theta_\gamma$ &
$\theta_\tau$ & $\theta_{\phi_1,0}$ &$\theta_{\phi_1,1}$ &
$\theta_{\phi_2,0}$ &$\theta_{\phi_2,1}$ \\
\hline
Estimate & $-4.90$ & 1.40 & 0.99 & 0.22 & $-0.05$ & $-0.58$ & 0.55 &
0.42\\
(S.E.) & (0.96) & (0.37) & (0.58) & (0.11) & (0.52) & (0.35) & (0.55) &
(0.28)\\
\hline
\end{tabular}

\footnotesize{Parameters shown are in unconstrained form, for example,
$\theta_\lambda = \log\lambda$.\\$\phi_1$
and $\phi_2$  are the scale
parameters for the treatment and control group,
respectively.}
\end{center}
\end{table}

  An interesting finding is that, on the basis of BIC, the model without
diabetes, the earlier Model 7, would be
preferred to Model 7(b), whereas the improvement upon accounting for
treatment was clear (compare Models 7 and 8). This
 reflects the fact that the difference between treatment groups is larger
that the difference within treatment groups (via the diabetes type). On
the other
 hand, based on the AIC and BIC, the models with a common diabetes effect
would not be chosen over the treatment-only model (actually,
Model
 7(c) does have a very slightly lower AIC than Model 7 but this is not
enough to warrant selection of the more complex Model 7(c)). In other
words, a
 common diabetes effect is not plausible. This is clear from Figure
\ref{fig:KMfitD} where the diabetes effect is reversed when comparing the
treatment group
 to the control group. Furthermore, we see from Table
\ref{tab:phimodD} --- which gives parameter estimates for Model 7(b)
--- that the diabetes effects within each group
($\theta_{\phi_1,1}$ and
 $\theta_{\phi_2,1}$, respectively) have different signs.\medskip

We can readily
 quantify the
treatment effect for those with juvenile and adult diabetes in terms of
quantile ratios. These are
 $\psi_J = \exp(\theta_{\phi_2,0} - \theta_{\phi_1,0})$ and $ \psi_A =
\exp\{(\theta_{\phi_2,0} + \theta_{\phi_2,1}) - ( \theta_{\phi_1,0} +
\theta_{\phi_1,1})\}.$ Their estimates turn out to be $\hat\psi_J = 1.81$,
with 95\% confidence interval $(1.12,2.91)$, and $\hat\psi_A = 4.94$, with
95\% confidence interval $(2.71,9.02)$, respectively. The
 conclusion is that time to blindness is almost doubled for treated
individuals with juvenile diabetes, while the time to blindness is
increased by a
 factor of almost five for treated individuals with adult diabetes.
(Similar point estimates of quantile ratios arise from HBS's model also.)
\medskip

\mage{Finally, we} \kevin{also investigate whether or not covariate dependent copula
parameters ($\lambda$
and $\omega$) might be needed. We therefore} \mage{extended} \kevin{Model 7(b) (of
Table
\ref{tab:phimodD}) to have covariate dependent copula parameters. The resulting model
(not shown) has AIC and BIC values of $1662.51$ and $1695.34$, respectively,
which are both higher than those of Model 7(b). Moreover, the estimated $K$ values for
$D=0$ and $D=1$ are respectively 0.19 and 0.20 which are numerically very close
to each other --- and indeed to that of Model 7(b). Furthermore, the fitted marginal
models (not shown) are almost indistinguishable from those seen in Figure
\ref{fig:KMfitD}. Therefore, here, covariate-dependent copula parameters are not
supported by the data,} \mage{that is,} \kevin{Model 7(b) is sufficiently flexible. Of
course, in
other settings the level of correlation may vary with covariates, and modelling
correlation on covariates in addition to the marginal distributions may avoid model
misspecification (again noting the} \mage{effect on Kendall's tau of the marginal
model as} \kevin{observed in Table \ref{tab:proftau}).
}

\medskip
\bigskip
\noindent \textbf{6.~Further remarks}\medskip\medskip

In this article, we have proposed the novel combination of APGW marginal
distributions and the BB9/PVF
copula. We have shown that this specific unification is very natural and effective,
yielding a new
bivariate model whose marginals include many of the most popular survival
distributions (and, indeed, whose marginals may differ in type via separate
$\tau$ parameters, as shown in (\ref{fullmodel})). This flexibility, along with the
variety of
regression
structures available, produces a very general overall modelling scheme which is useful
in practice.
\medskip

On the practical implementation of our  model, it is worth
highlighting that, based on the findings of BJN,
we would not require a model with both $\lambda_i$ and $\phi_i$
(a vertical scale parameter and a horizontal scale parameter,
respectively)
appearing in the APGW marginal survival function $S_{i,A}(z_i)$, $i=1,2$,
simultaneously, due to
their similar roles. However, here, with $\lambda_1 = \lambda_2 =
\lambda$, where
$\lambda$ also controls dependence within $\widehat{C}(u,v)$, we do not
experience
the issue. It is possible that estimation instability may arise within the
model when
$\phi_1=\phi_2=\phi$ particularly when $\omega$ is close to one as, in
that case, $\lambda$ plays little role in characterising dependence and
the margins then
simply contain two scale parameters, $\lambda$ and $\phi$. Of course,
 one could contemplate models where
$\lambda_1$
and $\lambda_2$ are unconstrained (that is, $\lambda_1,\lambda_2\ne
\lambda$) but, following BJN, we would then consider either
$\text{APGW}(\lambda_i,\phi_i=1,\gamma_i,\tau_i)$ or
$\text{APGW}(\lambda_i=1,\phi_i,\gamma_i,\tau_i)$ margins.
\bigskip

Shared frailty models like the ones of interest in this article are
sometimes criticised on grounds of insufficient flexibility. A correlated
frailty model is an attractive alternative, but in order to obtain one
in the current context it
is necessary to employ a defensible
bivariate tempered stable/power variance distribution for the frailties.
We are not aware of such a bivariate TS/PV model.
\medskip
\bigskip

\noindent \textbf{References}\medskip

\hangindent=0.3in
\hangafter=1
\red{
Bagdonavi\c{c}ius, V. \& Nikulin, M.  (2002) \textsl{Accelerated Life Model;
Modeling and Statistical Analysis}.
Chapman \& Hall/CRC.}\vspace{3pt}

\hangindent=0.3in
\hangafter=1
Burke, K., Jones, M.C. \& Noufaily, A. (2018) A flexible  parametric
 modelling framework for survival analysis. Submitted.\vspace{3pt}

\hangindent=0.3in
\hangafter=1
Clayton, D.G. (1978) A model for association in bivariate life tables and
its application in epidemiological studies of familial tendency in chronic
disease incidence. \textsl{Biometrika}, 65, 141--151.
\vspace{3pt}

\hangindent=0.3in
\hangafter=1
Crowder, M. (1989)
A multivariate distribution with Weibull connections.
\textsl{J.\ Roy.\ Statist.\ Soc.\ Ser.\ B}, 51, 93--107.
\vspace{3pt}

\hangindent=0.3in
\hangafter=1
\red{Dimitrakopoulou, T., Adamidis, K.\ \& Loukas, S. (2007)
A lifetime distribution with an upside-down bathtub-shaped hazard function.
\textsl{IEEE\ Trans.\ Reliab.}, 56, 308--311.}
\vspace{3pt}

\hangindent=0.3in
\hangafter=1
Duchateau, L. \& Janssen, P.  (2008) \textsl{The Frailty Model}.
Springer.\vspace{3pt}

\hangindent=0.3in
\hangafter=1
\mcj{Fischer, M.\ \& Jakob, K. (2016) ptas distributions with application
to risk management. \textit{J.\ Statist.\ Distributions Applic.}, 3,
1-18.}\vspace{3pt}

\hangindent=0.3in
\hangafter=1
\mcj{Gupta, P.\ \& Gupta, R. (1996) Ageing characteristics of the Weibull
mixtures. \textit{Prob.\ Eng.\ Info.\ Sci.}, 10, 591-600.}\vspace{3pt}

\hangindent=0.3in
\hangafter=1
Hougaard, P. (1986) Survival models for heterogeneous populations derived
from stable distributions. \textit{Biometrika}, 73, 387-396.\vspace{3pt}

\hangindent=0.3in
\hangafter=1
Hougaard, P. (2000) \textsl{Analysis of Multivariate Survival Data}.
Springer.\vspace{3pt}

\hangindent=0.3in
\hangafter=1
Huster, W.J., Brookmeyer, R. \& Self, S.G. (1989)
Modelling paired survival data with covariates. \textsl{Biometrics}, 45,
145--156.
\vspace{3pt}

\hangindent=0.3in
\hangafter=1
Joe, H.  (1997) \textsl{Multivariate Models and Dependence Concepts}.
Chapman \& Hall.\vspace{3pt}

\hangindent=0.3in
\hangafter=1
Joe, H.  (2014) \textsl{Dependence Modeling with Copulas}.
Chapman \& Hall.\vspace{3pt}

\hangindent=0.3in
\hangafter=1
\red{Jones, M.C.\ \& Noufaily, A. (2015)
Log-location-scale-log-concave distributions for survival and reliability
analysis. \textsl{Elec.\ J.\
Statist.}, 9, 2732--2750.}
\vspace{3pt}

\mage{
\hangindent=0.3in
\hangafter=1
Nelsen, R.B. (2006) \textsl{An Introduction to Copulas,} 2nd.\ ed.
Springer.
\vspace{3pt}}

\hangindent=0.3in
\hangafter=1
\red{Nikulin, M.\ \& Haghighi, F. (2009)
On the power generalized Weibull family: model for cancer censored data.
\textsl{Metron}, 67, 75--86.}
\vspace{3pt}

\hangindent=0.3in
\hangafter=1
Oakes, D. (1989)
Bivariate survival models induced by frailties. \textsl{J.\ Amer.\
Statist.\ Assoc.}, 84, 487--493.
\vspace{3pt}

\hangindent=0.3in
\hangafter=1
Romeo, J.S., Meyer, R.\ \& Gallardo, D.I. (2018)
Bayesian bivariate survival analysis using the power variance function
copula. \textsl{Lifetime Data Anal.}, 24, 355--383.
\vspace{3pt}

\hangindent=0.3in
\hangafter=1
\mcj{Tweedie, M. (1984) An index which distinguishes between some
important exponential families. In Ghosh, J.\ and Roy, J., editors,
\textsl{Statistics: Applications and New Directions}, pp.~579--604.}
\vspace{3pt}

\hangindent=0.3in
\hangafter=1
Wienke, P. (2011) \textsl{Frailty Models in Survival Analysis}.
Chapman \& Hall.

\bigskip
\bigskip

\noindent \textbf{\mcj{Appendix A: Frailty Link for APGW
Distribution}}\medskip

Write APGW$(\mcj{\gamma,\,}\kappa, \lambda)$ for the adapted PGW
distribution with
proportionality parameter $\lambda>0$ and shape parameter $\kappa$ i.e.\
with c.h.f.\ \(\lambda\, H_A(t^\gamma;\kappa) \). Also, write
the three-parameter version of the TS distribution as $TS(\omega,\xi,
\theta)$,  $0 \leq \omega \leq 1$,
$\xi>0,$ $\theta \geq 0$, having Laplace transform
\[{\cal L}^H_{\omega,\xi,\theta} (s) = \exp\left[-\frac\xi\omega
\left\{(\theta+s)^\omega -\theta^\omega \right\}\right]\]
and density $g^H_{\omega,\xi,\theta}$.\medskip

\noindent {\sc Result A1}.
Let $T|B=b \sim \textrm{APGW}(\mcj{\gamma,\,}\kappa, b)$, $\kappa>0$, and
let
\(B \sim\)\linebreak \(\textrm{TS}\left(\omega,
\frac{\kappa^{\omega-1}(\omega\kappa+1)}{(\kappa+1)^{\omega}},
\frac{\kappa+1}\kappa\right)\).
Then, $aT\sim \textrm{APGW}(\mcj{\gamma,\,}\omega \kappa, 1)$ where \(a
=\)\linebreak \(
\{(\kappa+1)/(\omega\kappa+1)\}^{1/\gamma}\).\medskip

\noindent {\tt Proof}
\begin{eqnarray*}
S_T(t)&=& \int_0^\infty \exp\{-bH_A(t^\gamma;\kappa) \}
g^{H}_{\omega,
\frac{\kappa^{\omega-1}(\omega\kappa+1)}{(\kappa+1)^{\omega}},
\frac{\kappa+1}\kappa}(b) \,db\\
&=& {\cal L}^H_{\omega,
\frac{\kappa^{\omega-1}(\omega\kappa+1)}{(\kappa+1)^{\omega}},
\frac{\kappa+1}\kappa}\{H_A(t^\gamma;\kappa)\}\\
&=&
\exp\left[-\frac{\kappa^{\omega-1}(\omega\kappa+1)}{\omega(\kappa+1)^\omega}
 \right.\\
&& \left. ~~~~~\times
\left\{\left(\frac{\kappa+1}\kappa + \frac{\kappa+1}\kappa
\left\{\left(1+\frac{t^\gamma}{\kappa+1}\right)^\kappa-1\right\}
\right)^{\omega}
- \left(\frac{\kappa+1}\kappa \right)^{\omega}
\right\}\right] \\
&=&
\exp\left[-\frac{(\omega\kappa+1)}{\omega\kappa}
\left\{\left(1+\frac{t^\gamma}{\kappa+1}\right)^{\omega\kappa}-1\right\}
\right]
= \exp\left[-H_A\left\{ \left(\frac{t}{a}\right)^\gamma;
\omega\kappa\right\}\right].\,\,\,\,\Box
\end{eqnarray*}\smallskip

\noindent \textbf{Appendix \mcj{B}: Proof of (\ref{bound})}\medskip

For the Gumbel copula, $S=12\, {\cal I}-3$ where
\begin{eqnarray*}
{\cal I} &=& \int_0^1\int_0^1 \widehat{C}_G(u,v) \,dv\,du\\
&=& \int_0^1\int_0^1 \exp\left[-\left\{(-\log u)^{1/\omega}+(-\log
v)^{1/\omega}\right\}^\omega \right]\,dv\,du\\
&=& \int_0^\infty\int_0^\infty e^{-(x+y)}
\exp\left\{-\left(x^{1/\omega}+y^{1/\omega}\right)^\omega \right\}\,dy\,dx
\\
&\leq & \int_0^\infty\int_0^\infty
\exp\left\{-(1+2^{\omega-1})(x+y)\right\} \,dy\,dx\\
&= & 1/(1+2^{\omega-1})^2,
\end{eqnarray*}
hence (\ref{bound}). Here, we have used the generalized mean inequality
$ (x^{1/\omega}+y^{1/\omega})^\omega \geq 2^{\omega-1}(x+y)$
for $0\leq\omega \leq 1$.

\medskip
\bigskip

\noindent \textbf{Appendix \mcj{C}: Formulae Underlying Claims in
Section 4.2}\medskip

From (\ref{LA}), we have
$$
L_A^{10}(z_1,z_2)= \frac{\gamma_1}{\omega} z_1^{\gamma_1-1}
\left(1+\frac{z_1^{\gamma_1}}{\tau_1+1}\right)^{\tau_1-1}
\left\{\frac{(\tau_1+1)}{\tau_1}
\left(1+\frac{z_1^{\gamma_1}}{\tau_1+1}\right)^{\tau_1}-\frac1{\tau_1}
\right\}^{(1/\omega)-1}
$$
and
\begin{eqnarray*}
L_A^{20}(z_1,z_2)&= & \frac{\gamma_1}{\omega\,\tau_1} z_1^{\gamma_1-2}
\left(1+\frac{z_1^{\gamma_1}}{\tau_1+1}\right)^{\tau_1-2}
\left\{\frac{(\tau_1+1)}{\tau_1}
\left(1+\frac{z_1^{\gamma_1}}{\tau_1+1}\right)^{\tau_1}-\frac1{\tau_1}
\right\}^{(1/\omega)-2} \\
&&~~~~~~~~~~\times \left[
\left\{(\gamma_1-1)(\tau_1+1)+\left(
\frac{\gamma_1\,\tau_1}\omega-1\right)
z_1^{\gamma_1} \right\}
\left(1+\frac{z_1^{\gamma_1}}{\tau_1+1}\right)^{\tau_1}\right.\\
&&\left.~~~~~~~~~~~~~~~~~~~~~~~- \left\{\gamma_1-1+
\frac{(\gamma_1\tau_1-1)}{\tau_1+1}z_1^{\gamma_1}\right\}\right].
\end{eqnarray*}
Using these formulae in place of the equivalent formulae for $L_N$ in
(\ref{hazgederiv}), we find that
$h'(z_1|Z_2 \geq z_2)$ is equal to positive terms
times
\begin{eqnarray}
&&
\left[\left\{\frac{(\tau_2+1)}{\tau_2}
\left(1+\frac{z_2^{\gamma_2}}{\tau_2+1}\right)^{\tau_2}
-\frac1{\tau_2}\right\}^{1/\omega} -1\right]\nonumber\\
&&~~~~~~~~~~\times \left[
\left\{(\gamma_1-1)(\tau_1+1)+\left(
\frac{\gamma_1\,\tau_1}\omega-1\right)
z_1^{\gamma_1} \right\}
\left(1+\frac{z_1^{\gamma_1}}{\tau_1+1}\right)^{\tau_1}\right.\nonumber\\
&&\left.~~~~~~~~~~~~~~~~~~~~~~~- \left\{\gamma_1-1+
\frac{(\gamma_1\tau_1-1)}{\tau_1+1}z_1^{\gamma_1}\right\}\right]\label{appa}\\
&&+
\left\{\frac{(\tau_1+1)}{\tau_1}
\left(1+\frac{z_1^{\gamma_1}}{\tau_1+1}\right)^{\tau_1}
-\frac1{\tau_1}\right\}^{1/\omega}\nonumber\\
&&~~~~~~~~~~\times \left[
\left\{(\gamma_1-1)(\tau_1+1)+\left(
{\gamma_1\,\tau_1}-1\right)
z_1^{\gamma_1} \right\}
\left(1+\frac{z_1^{\gamma_1}}{\tau_1+1}\right)^{\tau_1}\right.\nonumber\\
&&\left.~~~~~~~~~~~~~~~~~~~~~~~- \left\{\gamma_1-1+
\frac{(\gamma_1\tau_1-1)}{\tau_1+1}z_1^{\gamma_1}\right\}\right]\nonumber
\end{eqnarray}
from which the claims about the monotonicity properties of
$h(z_1|Z_2 \geq z_2)$ in Section 4.2 follow. Similarly, the
 term in square brackets in the formula (\ref{hazeqderiv}) when $L_A$
replaces $L_N$ is  equal to positive terms times a formula identical to
(\ref{appa}) except with the term `$\gamma_1\tau_1$' replaced by zero in
its
fifth line.
This sole difference is responsible for the  conclusions in Section 4.2
concerning the monotonicity  of
$h(z_1|Z_2 = z_2)$.

\end{document}